\documentclass [prb,twocolumn,superscriptaddress,showpacs,preprintnumbers] {revtex4}
\usepackage[]{amssymb,amsmath,amsfonts}
\usepackage[]{graphicx}

\begin{document}
\title{Ab-initio calculation of all-optical time-resolved calorimetry\\of nanosized systems: Evidence of nanosecond-decoupling of electron\\and phonon temperatures}

\author{F. Banfi} \email[]{francesco.banfi@dmf.unicatt.it}
\affiliation{Dipartimento di Matematica e Fisica, Universit\`a Cattolica, I-25121 Brescia, Italy}

\author{F. Pressacco}
\affiliation{Dipartimento di Fisica, Universit\`a degli Studi di Trieste and Sincrotrone Trieste, Basovizza, I-34012 Trieste, Italy}

\author{B. Revaz}
\affiliation{\'Ecole Polytechnique F\'ed\'erale de Lausanne, Laboratoire de Microsyst\`emes 3, CH-1015 Lausanne, Switzerland}

\author{C. Giannetti}
\affiliation{Dipartimento di Matematica e Fisica, Universit\`a Cattolica, I-25121 Brescia, Italy}

\author{D. Nardi}
\affiliation{Dipartimento di Matematica e Fisica, Universit\`a Cattolica, I-25121 Brescia, Italy}
\affiliation{Dipartimento di Fisica, Universit\`a degli Studi di Milano, I-20122 Milano, Italy}

\author{G. Ferrini}
\affiliation{Dipartimento di Matematica e Fisica, Universit\`a Cattolica, I-25121 Brescia, Italy}

\author{F. Parmigiani}
\affiliation{Dipartimento di Fisica, Universit\`a degli Studi di Trieste and Sincrotrone Trieste, Basovizza, I-34012 Trieste, Italy}

\date{\today}

\begin{abstract}
The thermal dynamics induced by ultrashort laser pulses in nanoscale systems, i.e. all-optical time-resolved nanocalorimetry is theoretically investigated from 300 to 1.5 K. We report ab-initio calculations describing the temperature dependence of the electron-phonon interactions for Cu nanodisks supported on Si. The electrons and phonons temperatures are found to decouple on the ns time scale at $\sim$ 10 K, which is two orders of magnitude in excess with respect to that found for standard low-temperature transport experiments. By accounting for the physics behind our results we suggest an alternative route for overhauling the present knowledge of the electron-phonon decoupling mechanism in nanoscale systems by replacing the mK temperature requirements of conventional experiments with experiments in the time-domain.
\end{abstract}

\maketitle
In the last decade the advent of nanoprocessing techniques has emphasized the need for designing non-invasive methodologies to access the realm of the thermal properties of nanoscale systems.\cite{Cahill2003,Schwab2000,Giazotto2006,Juve2009} However, conventional calorimetry is limited to samples of few tens of micrograms. A successful improvement recently arose from the micro-membrane-based nanocalorimeters.\cite{Schwab2000,Fon2005} These devices perform well, in terms of sensitivity, but they are limited to cryogenic temperatures and to time-integrated applications and need on-chip integration.\\
\indent When confronted with the problem of measuring the specific heat of nanoscale objects, a fast non-contact probe is an optimal choice, the speed requirement being dictated by the fact that the time for heat exchange between the sample and the thermal reservoir is proportional to the sample mass\cite{Giannetti2009} and the non-contact probe avoids the addendum heat capacitance contribution, i.e. the heat capacity of the probe itself.\\
\indent In these last years experiments have been reported, along with the all-optical schemes,\cite{Comin2006,Giannetti2007, Giannetti2009, Nardi2009, Siemens2009, Muskens2006} that unlocked the gate for all-optical time-resolved nanocalorimetry. However, the physical mechanism involved in an all-optical time-resolved nanocalorimetry are, over a wide temperature range, still unexplored. All-optical time-resolved nanocalorimetry consists of a system where a nanosample, a metallic nanodisk in the present work, is placed in thermal contact on a substrate serving as a thermal bath. An ultrafast laser pump beam delivers an energy density $\delta$$U_{V}$ to the nanodisk. The sample's temperature time relaxation to the substrate is measured via a time-delayed probe beam. Several detection schemes can be exploited or envisioned. Time Resolved Thermoreflectance (TR-TR) measures the temperature dependent changes in the optical reflectivity,\cite{Stoner1993,Cahill2003} Time-Resolved Spatial Modulation Spectroscopy (TR-SMS)\cite{Muskens2006} and time-resolved Near-Infrared Diffraction (TR-NIRD)\cite{Giannetti2007, Giannetti2009} detect the transmission and reflectivity changes modulated by the nano-object thermal expansion, whereas Time-Resolved X-ray diffraction (TR-XRD) reveals the transient lattice thermal expansion.\cite{Plech2003, Plech2004}\\
\indent In the present work the thermal dynamics occurring in time-resolved all-optical calorimetry is theoretically investigated from ambient to pumped liquid Helium temperatures. The theoretical frame accounts for temperature dependent material properties. In particular the microscopic electron-phonon interaction term $\Gamma$ is calculated within the frame of Allen theory,\cite{Allen1987PRL} starting from ab-initio Density Functional Theoretical calculations of the Eliashberg function. The nanodisk electrons $T_{el}$ and phonons $T_{ph}$ temperatures are found to decouple on the ns time scale at $\sim10$ K, that is two orders of magnitude in excess with respect to standard low-temperature transport experiments. The temperature decoupling extent, and the time scale over which it occurs, are ruled by the $\Gamma(T_{el},T_{ph})$ landscape. These findings set the limits of applicability of ultrafast nanocalorimetry well above 1.5 K while suggesting a new route to investigate the physics of electron-phonon interaction where the sub-K temperature requirement can be substituted by the ns time resolution.\\
\begin{figure}[t]
\centering
\includegraphics[bb=56 324 446 574,keepaspectratio,clip,width=0.98\columnwidth]{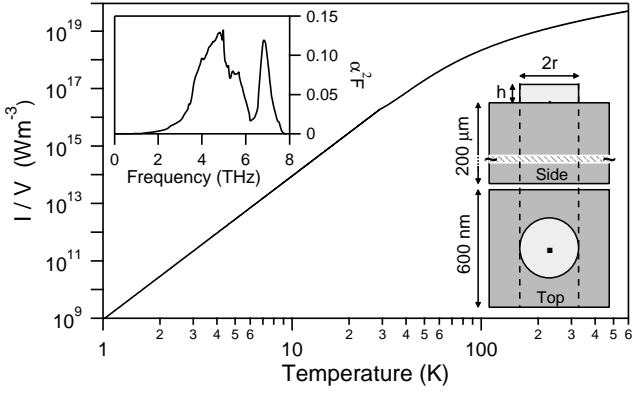}
\caption{Interaction term $I(T)/V$ in log-log scale obtained from numerical integration of Eq.~\ref{interaction}. Insets: Eliashberg function $\alpha^{2}F$ obtained from ab-initio Density Functional-based calculations (top left). Unit cell geometry (bottom right): $h$= 30 nm, $r$=150 nm.}
\label{Figure1_3}
\end{figure}
\indent The physics is here investigated considering a lattice of Cu nanodisks deposited on a Si substrate which unit cell dimensions are reported in inset of Fig.~\ref{Figure1_3}. The bottom of the Si substrate is kept at constant temperature $T_{cryo}$ by a cryostat, whereas insulating boundary conditions apply to the Si cell lateral boundaries. Assuming sample's excitation by a single Ti:sapphire pump pulse (120 fs pulse duration, 1 nJ per pulse, 800 nm central wavelength, 40 $\mu$m spatial extension at FWHM, and 1 MHz repetition rate), the thermal evolution problem can be conveniently envisaged by a three steps mechanism. In the first step, the laser pulse heats the electron gas of the metallic nanodisks (sub-ps time scale). In the second step, the hot electron gas thermalizes with the phonons within the disk (ps time scale). The physics entailed in the first two steps is well modeled by the Two Temperatures Model:\cite{Kaganov1957}
\begin{eqnarray}
&&C_{el}(T_{el})\partial_{t}T_{el}=P_{p}(t)-\Gamma(T_{el},T_{ph})+\vec{\nabla}\cdot(k_{el}(T_{el})\vec{\nabla}T_{el})\nonumber\\
&&C_{ph}(T_{ph})\partial_{t}T_{ph}=\Gamma(T_{el},T_{ph})+\vec{\nabla}\cdot(k_{ph}(T_{ph})\vec{\nabla}T_{ph}) \label{TTM}
\end{eqnarray} 
where $T$, $k$ and $C$ indicate the temperature, thermal diffusion coefficient and specific heat per unit volume, respectively, the reference to the electrons ($el$) or Cu phonons ($ph$) being indicated by the subscript, and $P_{p}$ is the profile of the pulsed power per unit volume absorbed by the sample. The energy density absorbed by the sample is peaked within the nanodisk, because of the difference in the optical penetration depth between Cu and Si. This occurrence gives rise, in the third step, to the onset of an heat flux from the nanodisk to the substrate (ns time scale). The thermal link translates in the following boundary conditions at the disk-substrate interface:
\begin{eqnarray} 
\widehat{n}_{ph}\cdot k_{ph}\vec{\nabla}T_{ph}+(T_{ph}-T_{Si})/\rho_{th}(T_{ph})=0\nonumber\\
-\widehat{n}_{Si}\cdot k_{Si}\vec{\nabla}T_{Si}-(T_{ph}-T_{Si})/\rho_{th}(T_{ph})=0 \label{boundary_resistivity}
\end{eqnarray} 
$\widehat{n}_{ph}$ and $\widehat{n}_{Si}$ being the outward unit vector normal to the disk and Si boundary, respectively, and $\rho_{th}$ the temperature dependent boundary thermal resistivity. The temperature within the Si substrate is calculated via the standard Fourier heat transfer equation. This three-steps sequence repeats itself upon arrival of a new laser pulse, once every $1\mu$s. The steady-state temperature distribution, due to the pulses train, is modeled following Ref.~\onlinecite{Giannetti2007}. This distribution serves as the initial boundary condition for the thermal dynamics following the arrival of a single pulse and is rather constant within the first few microns region beneath the Si-nanodisk interface. In the following the temperature in this region will be addressed as $T_{Si}$, the temperature calculated at a point 5 nm beneath the disk-Si interface [inset of Fig.~\ref{Figure2_3}(a)]. The temperature-dependent specific heats and thermal conductivities entering the equations have been taken from data available in the literature.\cite{note1}\\
\begin{figure}[t]
\centering
\includegraphics[bb=76 120 464 480,keepaspectratio,clip,width=0.98\columnwidth]{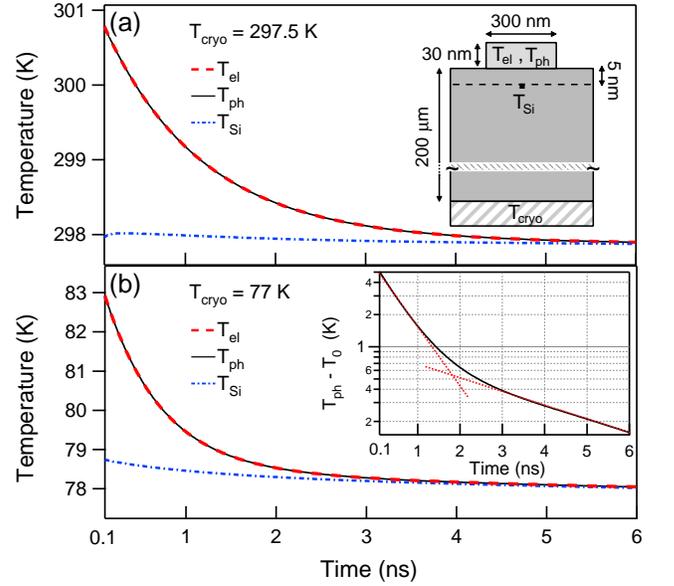}
\caption{(Color online) Simulations results for $T_{cryo}$=297.5 and 77 K. Insets. Panel (a): locations within the sample were the temperatures are calculated. Panel (b): the difference between $T_{ph}$ and its asymptotic value $T_{0}$, in semi-log scale; dashed red lines are a guide to the eye, showing double exponential behavior.}
\label{Figure2_3}
\end{figure}
\indent A fundamental issue is the temperature dependence of the electron-phonon coupling which, in the high temperature limit, reads $\Gamma=G(T_{el}-T_{ph})$. This approximation fails when the thermal dynamics spans the entire range from hundreds of degrees K to liquid Helium temperature. For instance, with $T_{cryo}$ set at 4.2 K, $T_{el}$ rises to $\sim100$~K within the pump pulse time width, relaxing back to $\sim$ 4.2 K on the ns time scale. In order to properly account for the temperature dependence over such a wide temperature range, the electron-phonon coupling is cast as $\Gamma=[I(T_{el})-I(T_{ph})]/V$ where 
\begin{equation}
I(T)=2\pi N_{c}N_{E_{F}}\int_{0}^{\infty}d\omega\alpha^{2}F(\omega)(\hbar\omega)^{2}n_{BE}(\omega,T)
\label{interaction}
\end{equation} 
with $N_{c}$ the number of cells in the sample, $N_{E_{F}}$ the electronic DOS at the Fermi level, $n_{BE}$ the Bose-Einstein distribution and $\alpha^{2}F$ the Eliashberg function,\cite{Allen1987PRL} here calculated ab-initio within the frame of Density-Functional Theory.\cite{note2} Simulations results are reliable down to 0.6 THz, the electron-phonon interaction calculation failing for longer phonon wavelengths. For lower frequencies $\alpha^{2}F=\widetilde{\lambda}\omega^{2}/\omega^{2}_{D}$,\cite{Allen1987PRL} $\omega_{D}$ being the Debye frequency and $\widetilde{\lambda}$ a fitting parameter. The value $\widetilde{\lambda}$ has been set to have $I(T)$ matching the experimental data reported at sub-K temperatures.\cite{Giazotto2006} The coefficient $G$ occurring in the expression for the high temperature electron-phonon interaction, obtained from the linear fit of $I(T)/V$ for $T\geqslant \Theta_{D}$, with $\Theta_D$ the Debye temperature, is $8.43\cdot10^{16}$ W/(m$^{3}$K), in good agreement with experimental values reported in the literature.\cite{Brorson1990, Elsayed1987} Results for both $I(T)/V$ and $\alpha^{2}F$ are shown in Fig.~\ref{Figure1_3}. Attention is drawn on the nine orders of magnitude change of the interaction term in the temperature range of interest for this work, i.e. from ambient temperature to 1.5 K.\\
\indent The thermal boundary resistivity is modeled by the Acoustic Mismatch Model (AMM): $\rho_{th}(T)=A_{bd}T^{-3}$ for $T\leqslant\widetilde{T}=30$ K with $A_{bd}=1.4\cdot10^{-3}$ K$^{4}$m$^{2}$W$^{-1}$ specific for a Cu-Si interface.\cite{Swartz1987APL} For higher temperatures, $\rho_{th}$ is taken at the constant value $A_{bd}\widetilde{T}^{-3}$, that is within the range of values reported for ambient temperature. Nevertheless, to dissipate all doubts regarding the dependence of our findings on the detailed value of $\widetilde{T}$, calculations have been performed with values of $\widetilde{T}$ spanning the range 20-50 K without affecting the physics.\\
%
\indent The energy density delivered to the nanodisk reads $\delta U_{V}$=$\delta Q_{V}+ \delta W_{V}$, where $\delta Q_{V}$ and $\delta W_{V}$ are the thermal and mechanical energy density respectively. In the present theoretical description only the thermal channel has been considered: $\delta U_{V} \simeq \delta Q_{V}=C_{ph} \delta T_{ph}$, hence yielding a temperature increase $\delta T_{ph} \simeq \delta U_{V}/C_{ph}$. In order to justify the above mentioned approximation we calculate the ratio of the mechanical energy correction to the thermal energy term. The temperature increase $\delta T_{ph}$ triggers a thermal expansion $\delta r/r=\alpha \delta T_{ph}$,  $\delta r/r$ being the disk's strain and $\alpha$ the linear thermal expansion coefficient. Associated with the disk thermal expansion is an increase in mechanical energy density.  We estimate the system's mechanical energy density assuming it concentrated in a nanodisk radial breathing mode, $\delta W_{V}=(Y/2)(\delta r/r)^{2}=(Y/2)(\alpha \delta T_{ph})^{2}$, $Y$ being the nanodisk Young Modulus. The sought ratio reads $\delta W_{V}/\delta Q_{V}=(Y/2)(\alpha^{2}/C_{ph})\delta T_{ph}$$\sim$ 10$^{-5}$, as obtained upon substitution of the ambient temperature values \ $Y$=1.2$\times$10$^{11}$ Pa,\cite{HandbookChemistryPhysics} $\alpha$= 1.6$\times$10$^{-5}$ K$^{-1}$,\cite{HandbookOpticalMaterials} and $\delta T_{ph}\simeq3$~K as calculated following electron-phonon thermalization on the ps time scale. A mechanical energy correction five order of magnitudes smaller with respect to the thermal channel justifies our initial assumption.\\
\indent The focus of the present paper is on the temperature relaxation dynamics relevant for nanocalorimetry, therefore only the time scale from 100 ps to 10 ns will be discussed. Simulations results at ambient and liquid Nitrogen temperatures are reported in Fig.~\ref{Figure2_3}. In both cases the temperature within the nanodisk is well defined ($T_{ph}$=$T_{el}$) and remains homogeneous throughout its volume, thus assuring thermodynamical equilibrium between electrons and phonons and, consequently, the technique's applicability to investigate the nanodisk thermal dynamics.\\
\begin{figure*}[t]
\centering
\includegraphics[bb=104 96 528 332,keepaspectratio,clip,width=1.4\columnwidth]{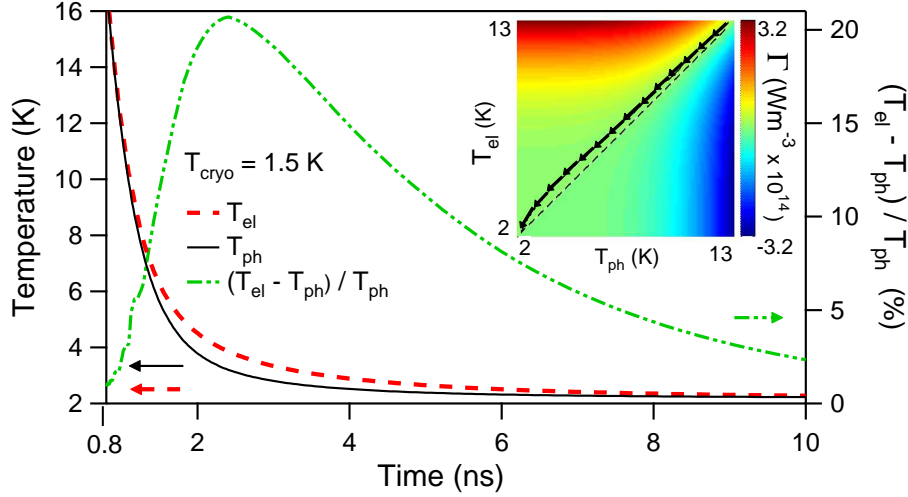}
\caption{(Color online) Simulations results for $T_{cryo}=1.5$~K. Left axis: $T_{ph}$ and $T_{el}$. Right axis: relative temperature variation in percentage. Inset: nanodisk trajectory in phase space for $T<13$ K (arrows) and $T_{el}$=$T_{ph}$ reference line (dashed) superposed on the interaction landscape $\Gamma$.}
\label{Figure3}
\end{figure*}
\indent At $T_{cryo}$=297.5 K the Si substrate acts as a thermal reservoir at constant temperature, while the disk's temperature time dependence follows a single exponential decay with time constant $\tau_{1}=1.1$ ns [Fig.~\ref{Figure2_3}(a)]. The maximum relative change of the nanodisk specific heat, $\Delta C_{ph}/C_{ph}$=1.14$\cdot10^{-3}$, allows to ignore its temperature dependence. These numerical results suggest modeling the problem as an isothermal disk, with initial temperature 301 K, thermally linked with a reservoir at $T_{Si}=298$~K. A value of the Biot number $Bi=h/k_{ph}\rho_{th}\sim10^{-3}$ guarantees the disk remains isothermal throughout the thermal relaxation process. Under these circumstances the analytic solution for the disk temperature follows a single exponential decay\cite{Necati} with time constant $\tau=1$ ns, in agreement with the numerical simulation, and the nanodisk specific heat is readily accessible as $\tau/h\rho_{th}$.\\
\indent At $T_{cryo}=77$ K, the nanodisk temperature dynamics, obtained from the numerical simulation, follows a double exponential with decay times $\tau_{1}=0.617$~ns and $\tau_{2}=6.286$~ns, respectively [Fig.~\ref{Figure2_3}(b)]. On the contrary to the previous case, $T_{Si}$ is not constant, due to the diminishing of C$_{Si}$ with temperature. The physics can be rationalized as follows: the isothermal nanodisk "feels" a substrate constant temperature on the sub-ps time scale and it thermalizes with it with $\tau=0.622$~ns, as calculated on the basis of the isothermal disk model exploited above. This interpretation is supported by the agreement between the model-calculated $\tau$ and the value $\tau_{1}$ from numerical simulations. On the longer time scale the disk and the Si substrate portion in close proximity to the disk-Si interface jointly thermalize with the rest of the Si substrate.\\
\indent The physics changes drastically for $T_{cryo}$ in the range of liquid He temperatures. Simulations results for T$_{cryo}$=1.5~$K$ are reported in Fig.~\ref{Figure3}. $T_{el}$ and $T_{ph}$ decouple around 13~K, 1~ns after the pump pulse arrival, with a maximum relative temperature variation of $22\%$, for a time delay of 2.2 ns. A similar result, with relative temperature variation of $6\%$, is found for $T_{cryo}=4.2$~K. From transport measurements in mesoscopic structures the electron-phonon temperature decoupling occurs at sub-K temperatures,\cite{Giazotto2006,Roukes1985} whereas for the present system our model foresees the decoupling at temperatures about two orders of magnitude higher. Interestingly here is to point out that the electron-phonon decoupling predicted in the present work, cannot be observed by transport measurements because of the lack of time resolution, rising a strong demand for ultrafast time-resolved nanocalorimetry. A fast version of Normal Superconductor Insulator thermometer actually achieved sub-$\mu$s read out resolution, thus, foreseeing the possibility to access thermal relaxations rates with transport measurements in the $\mu$s time-window. Nevertheless the time-window over which the decoupling here addressed occurs is on the ns-10 ns time scale, that is two orders of magnitude faster.\\
\indent The physics entailed in the calculations is conveniently unfolded as follows. The substrate temperature beneath the nanodisk reaches its asymptotic value of 2.2~K for a time delay of 1 ns. The system thermal dynamics is then well described in a Two Dimensional (2D) phase space of coordinates ($T_{ph}$, $T_{el}$) by the following set of equations:
\begin{eqnarray} 
&&\partial_{t}T_{el}=-\Gamma(T_{el},T_{ph})/C_{el}(T_{el}) \label{TTM_low_T}\\
&&\partial_{t}T_{ph}=-\Omega(T_{ph})/C_{ph}(T_{ph})+\Gamma(T_{el},T_{ph})/C_{ph}(T_{ph}) \nonumber
\end{eqnarray}
where the thermal flux per unit volume to the Si slab is taken into account by $\Omega(T_{ph})$=$(T_{ph}-2.2 K)/h\rho_{th}(T_{ph})$. The thermal conductivities within the disk have been omitted, the simulation results showing that the temperature distribution within the disk is spatially uniform over the ns time scale, for both electrons and phonons temperatures. Eq.~\ref{TTM_low_T} represents the velocity component in the 2D phase space, the initial conditions being the temperatures ($T_{ph}$, $T_{el}$) reached for a 1 ns delay. The nanodisk trajectory in phase space is tangent to the velocity field, $(\partial_{t}T_{ph},\partial_{t}T_{el})$. The trajectory is the line $T_{ph}=T_{el}$ for time delays spanning the ps to 1 ns range, the velocity vector pointing along this direction. At 13~K, $\partial_{t}T_{ph}>\partial_{t}T_{el}$ and the trajectory changes accordingly, hence $T_{ph}<T_{el}$. The velocity field is ruled by the interplay among $\Omega$, the specific heats and $\Gamma$. The $\Gamma$ interaction term is large enough in proximity of the line $T_{ph}=T_{el}$ to keep $T_{ph}$ and $T_{el}$ anchored down to 13~K, not so between 13 and 2.2~K. The system trajectory in phase space for $T<13$~K is reported, superimposed on the landscape $\Gamma$, in the inset of Fig. \ref{Figure3}.\\
\indent When compared with the results obtained for Cu nanodisks, the values for $\Omega$, $\Gamma$ and the specific heats for several metals in the temperature range were the decoupling is here shown to take place, suggest the present finding should occur in a wider range of materials other than Cu. To this aim an approximate analytic approach is here proposed, also serving as a valuable tool to further highlight the physical quantities ruling the electron-phonon temperature decoupling. In the low temperature limit, $T\ll\Theta_{D}$, well satisfied in the temperature range were the decoupling occurs, the interaction term approximates to $I(T)=(\Sigma_{0}/{V})T^{5}$, $\Sigma_{0}/V$ being the sub-K electron-phonon coupling constant.\cite{Giazotto2006} The Debye model for $C_{ph}$ applies, and Eq. \ref{TTM_low_T} reads:
\begin{eqnarray} 
&&\partial_{t}T_{el}=-C_{1}(\frac{\Sigma_{0}}{V})\frac{1}{N_{E_{F}}}(\frac{T_{el}^5-T_{ph}^5}{T_{el}})\\
&&\partial_{t}T_{ph}=-C_{2}\frac{\Theta_{D}^3}{nA_{bd}}(T_{ph}-T_{Si})+C_{3}(\frac{\Sigma_{0}}{V})\frac{\Theta_{D}^3}{n}(\frac{T_{el}^5-T_{ph}^5}{T_{ph}^3}) \nonumber\\
\label{appx_TTM_low_T}
\end{eqnarray}
with $C_{1}$, $C_{2}$ and $C_{3}$ positive material-independent constants, and $n$ the number of ions per unit volume of the chosen metal. In the case of a Cu nanodisk, $\partial_{t}T_{ph}<0$ at all times, thus $\partial_{t}T_{ph}$ is ruled by the power density delivered from the nanodisk's phonons to the substrate rather than the power density input from the electron to the phonon gas, hence the first term of the sum in Eq. \ref{appx_TTM_low_T} dominates over the second term.\\
\indent Let's now consider Al nanodisks instead and compare it to the Cu nanodisks case. With reference to Eq. \ref{appx_TTM_low_T}, the following ratios are calculated:
\begin{eqnarray} 
&&\nonumber(\partial_{t}T_{el})_{Al}/(\partial_{t}T_{el})_{Cu}\sim0.05\\
&&\nonumber(\frac{\Theta_{D}^3}{nA_{bd}})_{Al}/(\frac{\Theta_{D}^3}{nA_{bd}})_{Cu}\sim2.37\\
&&\nonumber(\frac{\Sigma_{0}}{V}\frac{\Theta_{D}^3}{n})_{Al}/(\frac{\Sigma_{0}}{V}\frac{\Theta_{D}^3}{n})_{Cu}\sim0.2
\label{aluminum}
\end{eqnarray}
%
The first ratio signifies that $T_{e}$ remains rather constant in the Al nanodisk as compared to the Cu case, whereas the last two ratios imply that the main contribution to $\partial_{t}T_{ph}$, also for the Al case, is the power density flow from the phonon gas to the Si substrate and that $T_{ph}$ relaxes to the substrate temperature two times faster than in the case of Cu nanodisks. The electron-phonon temperature decoupling is therefore expected to be more drastic in Al nanodisks as compared to the Cu case, the main role being played by $\Theta_{D}$ and the material-dependent constants entering $\Sigma_{0}$.\\
\indent Care is to be taken when addressing the system dimensions issue. In order to perform nanocalorimetry the condition $Bi<1$ needs to be fulfilled. This requirement poses constraints on the maximum disk height in relation to the adopted metals, i.e. pure metals versus metals alloys. For instance, investigating a Permalloy nanodisk at 300 K ($k_{Py}$=20 WK$^{-1}$m$^{-1}$ and $\rho_{th}$$\sim$10$^{-8}$ W$^{-1}$m$^{2}$K),\cite{MetalsHandbook, Swartz1987APL} $Bi\sim$1 for $h$=200 nm, whereas for a Cu nanodisk $Bi\sim1$ for $h=3.81$ $\mu$m. Furthermore, for low $k$ materials the laser penetration depth to disk height ratio can be an issue with regards to the disk temperature homogeneity. The nanoscale range guarantees for an isothermal nanodisk over a wide range of metals samples. On the other side, when dealing with metals nano-clusters, the low frequency vibrational eigenmodes arising from the cluster finite-size, 0.1 THz and below, should be accounted for in the Eliashberg function calculation.\\
\indent In conclusion, by modeling ab-initio the thermal dynamics induced by ultrashort laser pulses in nanoscale systems we show that the break-down of thermodynamical equilibrium between electrons and phonons on the ns time scale takes place at $\sim$10 K. This temperature is two orders of magnitude higher then that observed in standard transport measurements. These findings set the limits of applicability of ultrafast nanocalorimetry well above liquid He temperature, the electron-phonon temperature decoupling preventing a proper definition of the temperature concept of the nanosample as a whole. Finally, the present work, while making available a proper tool for interpreting all-optical time-resolved nanocalorimetry experiments, suggests a new route for investigating the physics of the electron phonon decoupling where the sub-Kelvin temperature requirement is substituted by the ns time resolution. If confirmed by the experiments, this discovery will bridge the fields of ultrafast optics and cryogenic transport in mesoscopic systems, while setting the investigation of the thermodynamics at the nanoscale into a new perspective.
\bibliography{Banfi_Biblio}

\begin{thebibliography}{26}
\bibitem[{not({\natexlab{b}})}]{note2}
\bibinfo{note}{The function $\alpha^{2}F$ and the DOS for Cu have been
  calculated implementing the code PWscf part of the Quantum ESPRESSO program.}
\end{thebibliography}
\end{document}